\documentclass[final,5p,times,twocolumn]{elsarticle}
\usepackage{amsmath}
\usepackage{lineno,hyperref}
\usepackage{notes2bib}
\usepackage[version=3]{mhchem} 
\modulolinenumbers[5]

\begin{document}

\begin{frontmatter}

\title{\texttt{LayerOptics}: Microscopic modeling of optical coefficients in layered materials }
\author{Christian Vorwerk}
\ead{christian.vorwerk@physik.hu-berlin.de}
\author{Caterina Cocchi}
\ead{caterina.cocchi@physik.hu-berlin.de}
\author{Claudia Draxl}
\address{Institut f\"ur Physik and IRIS Adlershof, Humboldt-Universit\"at zu Berlin, Berlin, Germany}
\address{European Theoretical Spectroscopic Facility (ETSF)}

\begin{abstract}
Theoretical spectroscopy is a powerful tool to describe and predict optical properties of materials. 
While nowadays routinely performed, first-principles calculations only provide bulk dielectric tensors in Cartesian coordinates.
These outputs are hardly comparable with experimental data, which are typically given by macroscopic quantities, crucially depending on the laboratory setup.
Even more serious discrepancies can arise for anisotropic materials, e.g., organic crystals, where off-diagonal elements of the dielectric tensor can significantly contribute to the spectral features.
Here, we present \texttt{LayerOptics}, a versatile and user-friendly implementation, based on the solution of the Maxwell's equations for anisotropic materials, to compute optical coefficients in anisotropic layered materials.
We apply this tool for post-processing full dielectric tensors of molecular materials, including excitonic effects, as computed from many-body perturbation theory using the \texttt{exciting} code. 
For prototypical examples, ranging from optical to X-ray frequencies, we show the importance of combining accurate \textit{ab initio} methods to obtain dielectric tensors, with the solution of the Maxwell's equations to compute optical coefficients accounting for optical anisotropy of layered systems.
Good agreement with experimental data supports the potential of our approach, in view of achieving microscopic understanding of spectroscopic properties in complex materials.
\end{abstract}


\end{frontmatter}

\section{Introduction}
First-principles methods represent a powerful tool to predict optical properties of materials with high accuracy.
In combination with experimental data, they provide insight into the features of the investigated systems. 
In solid-state physics, many-body perturbation theory (MBPT) is the state-of-the-art approach to compute dielectric properties \cite{onid+02rmp}: electron-electron correlation is included through the $GW$ approximation \cite{hedi65pr,hybe-loui85prl}, while excitonic effects are taken into account through the solution of the Bethe-Salpeter equation (BSE) \cite{hank-sham80prb,stri88rnc}.
Implementations of MBPT are now available interfaced to the most popular density-functional theory (DFT) packages, making these calculations routinely done. 
However, comparison with experimental data is often not straightforward.
Dielectric tensors computed from \textit{ab initio} codes are bulk quantities, relating to the coordinate system of the unit cell. 
This hardly fits typical laboratory conditions, where samples are usually thin films, including one or more layers of materials and a dielectric substrate \cite{sass+03sm,loi+05natm,raim+13jpcc,he+14cm}.
In anisotropic materials, where dielectric tensors have non negligible off-diagonal components, some spectral features can be completely missed, if only the diagonal terms are considered.
Finally, for a comparison with experimental data, calculations should take into account additional degrees of freedom, such as incidence angle and polarization of the incoming beam, as well as orientation and thickness of the sample. 

In this paper, we present \texttt{LayerOptics}, an efficient computational tool, based on the solution of Maxwell's equations for optically anisotropic media \cite{Yeh}, to compute Fresnel coefficients in layered materials.
This formalism is a generalization for anisotropic media of the 2$\times$2 approach, commonly used for isotropic layered systems \cite{2x2}.  
In the current implementation, \texttt{LayerOptics} is a post-processing tool for dielectric tensors obtained with the all-electron code \texttt{exciting} \cite{exciting}.
Due to its simple and versatile structure, interfacing \texttt{LayerOptics} to other \textit{ab initio} codes is straightforward.
After introducing the theoretical background and the structure of the implementation, we present the capabilities of \texttt{LayerOptics} with a selection of examples concerning optical and X-ray absorption properties of molecular materials, such as oligothiophene crystals and azobenzene self-assembled monolayers (SAMs).
Our results indicate the importance of off-diagonal elements of dielectric tensors in calculating optical properties of anisotropic thin films.
Significant changes in the Fresnel coefficients are observed by varying the angle of incidence of the incoming light as well as its polarization. 
We reproduce the spectra of a model organic crystal, including two layers of materials with different orientation of the molecules with respect to the substrate, as expected in experimental growth conditions.
Finally, we show how \texttt{LayerOptics} can be used to determine the parameters related to the orientation of the molecules in a SAM. 
Good agreement with experimental data supports the validity of our approach to reproduce optical absorption features in anisotropic layered materials.

The paper is organized as follows: In Section \ref{sec:theory} we provide the theoretical background for the calculation of the Fresnel coefficients. 
In Section \ref{sec:implement} we describe the adopted numerical procedure, and finally in Section \ref{sec:results} we present the application of \texttt{LayerOptics} to selected examples. 


\section{Theoretical Background}\label{sec:theory}
\begin{figure}
\centering
\includegraphics[width=.3\textwidth]{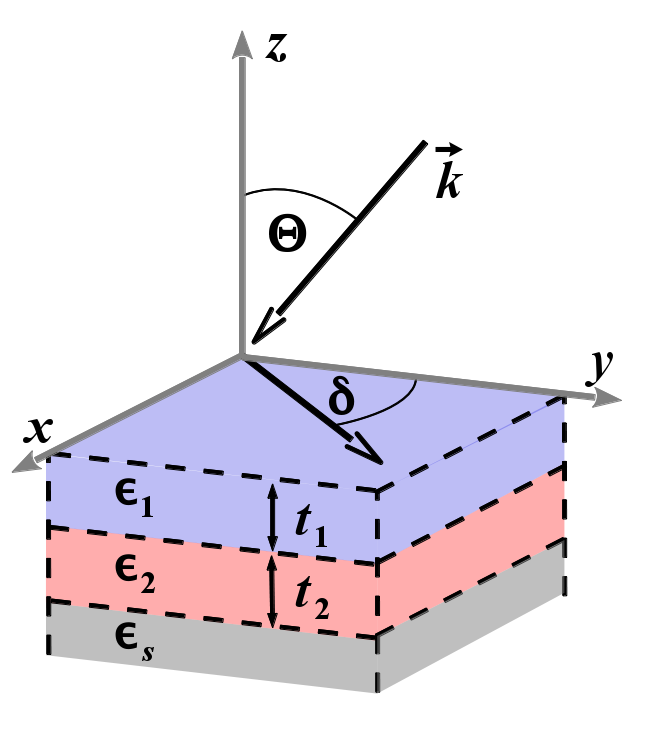}%
\caption{Schematic setup of a layered system on an isotropic substrate with dielectric tensor $\epsilon_S$. Each layer of material is characterized by its thickness ($t_1$ and $t_2$) and dielectric tensor ($\epsilon_1$ and $\epsilon_2$). $\mathbf{k}$ is the wave vector of the incoming light, with incidence angle $\Theta$, defined with respect to the surface normal $z$. The angle of polarization of the light in the medium ($\delta$) is indicated with respect to the $y$ axis.}
\label{fig1:setup}
\end{figure}

\subsection{Matrix formulation of Maxwell's equations}
The propagation of electromagnetic waves in an anisotropic material is determined by Maxwell's equation in momentum space:
\begin{equation}\label{eq1_1}
 \mathbf{k}\times\left( \mathbf{k}\times \mathbf{E}\right)+\frac{\omega^2}{c^2}\epsilon(\omega)\mathbf{E}=0 ,
\end{equation}
where $\mathbf{k}$ is the wave vector with frequency $\omega$, $c$ the velocity of light in vacuum, $\mathbf{E}$ the electric field and $\epsilon(\omega)$ the frequency-dependent dielectric tensor.
In order to obtain non trivial solutions for $\mathbf{E}$, the determinant of the homogeneous and linear system of equations (\ref{eq1_1}) has to vanish.
With fixed components $k_{x}$ and $k_{y}$, this condition yields four roots $k_{z,\sigma}$ ($\sigma = 1, ..., 4$), for each $\omega$. 
In well-behaved cases they correspond to two polarizations in two propagation directions ($\mathbf{p}_{\sigma}$) \bibnote{This is not necessarily the case on strongly anisotropic materials. See Appendix A.} .
We can write the total electric field in the medium as:
\begin{equation}\label{eq1_2}
 \mathbf{E}=\sum_{\sigma=1}^{4}A_{\sigma}\mathbf{p}_{\sigma}\exp \left[k_xx+k_yy+k_{z,\sigma}z-\omega t \right] ,
\end{equation}
and the corresponding magnetic-field vector 
as:
\begin{equation}\label{eq1_3}
\mathbf{H}=\frac{1}{\mu_0c}\mathbf{k}\times \mathbf{E} .
\end{equation} 
Transmission and reflection coefficients are obtained by imposing the boundary conditions of the parallel components of the electric (magnetic) field $E_x$ and $E_y$ ($H_x$ and $H_y$) at the layer interfaces. 
We assume the layered system to be infinitely extended in the $xy$-plane and stacked along the $z$ direction, as sketched in Fig. \ref{fig1:setup}.
The boundary between the top layer, which is by default a semi-infinite vacuum layer, and the 
first material layer is set at $z=0$. 
Each layer $n$, characterized by a dielectric tensor $\epsilon_n$, has finite thickness $t_{n}=z_{n-1}-z_{n}$, with $n = 1, ..., N$, where $N$ is the total number of layers.
A semi-infinite isotropic substrate, $S$, is assumed as bottom layer (see Fig. \ref{fig1:setup}).
By adopting these conventions, we can write the total dielectric tensor of the layered system as:
\begin{equation}\label{eq0_1}
 \epsilon= \left\{ \begin{array}{cc}
								\epsilon(0) &  z>0 \\
 								\epsilon(1) & 0>z>z_1 \\
 								\vdots\\
 								\epsilon(n) & z_{n-1}>z>z_n \\
 								\vdots \\
 								\epsilon(N) & z_{N-1}>z>z_N \\
 								\epsilon(S) & z_N>z .
 								
 \end{array}\right.
\end{equation}
At the interface $z=z_{n-1}$ this yields the matrix equation for the electric amplitudes:
\begin{equation}\label{eq1_4}
\left(\begin{array}{c}A_1(n-1)\\ A_2(n-1) \\ A_3(n-1) \\ A_4(n-1) \end{array}\right)= \mathbb{D}^{-1}(n-1)\mathbb{D}(n)\mathbb{P}(n)\left(\begin{array}{c}A_1(n)\\ A_2(n) \\ A_3(n) \\ A_4(n) \end{array}\right) ,
\end{equation} 
where $\mathbb{D}(n)$ and $\mathbb{P}(n)$ are 4$\times$4 matrices. 
The matrix $\mathbb{D}(n)$ includes the electric and magnetic polarization vectors $\mathbf{p}_{\sigma}(n)$ and $\mathbf{q}_{\sigma}(n)$: 
\begin{equation}\label{eq1_5}
\mathbb{D}(n)=\left( \begin{array}{cccc} p_{x,1}(n) & p_{x,2}(n) & p_{x,3}(n) & p_{x,4}(n)\\
										 q_{y,1}(n) & q_{y,2}(n) & q_{y,3}(n) & q_{y,4}(n)\\
										 p_{y,1}(n) & p_{y,2}(n) & p_{y,3}(n) & p_{y,4}(n)\\
										 q_{x,1}(n) & q_{x,2}(n) & q_{x,3}(n) & q_{x,4}(n)

\end{array}\right) ,
\end{equation}
while the matrix $\mathbb{P}(n)$ is formed directly from the $k_{z,\sigma}$-component:
\begin{equation}\label{eq1_6}
\mathbb{P}(n)=\left( \begin{array}{cccc} e^{ik_{z,1}(n)t_{n}} & 0 & 0 & 0\\
										 0 & e^{ik_{z,2}(n)t_{n}} & 0 & 0\\
										 0 & 0 & e^{ik_{z,3}(n)t_{n}} & 0\\
										 0 & 0 & 0 & e^{ik_{z,4}(n)t_{n}}\\

\end{array}\right).
\end{equation}
These relations between the electric field vectors at each layer boundary can be used to connect the amplitudes of the electric fields in the vacuum layer and in the substrate as follows:
\begin{equation}\label{eq1_7}
\mathbf{A}(0)=\mathbb{D}^{-1}(0)\mathbb{D}(1)\mathbb{P}(1)\mathbb{D}^{-1}(2)\mathbb{P}(1)\dots\mathbb{D}^{-1}(N)\mathbb{D}(S)\mathbf{A}(S) .
\end{equation} 
Denoting the total transfer matrix $\mathbb{T}$ as the product of the single-layer transfer matrices $\mathbb{T}(n)$:
\begin{equation}\label{eq1_8}
\mathbb{T}(n)=\mathbb{D}(n)\mathbb{P}(n)\mathbb{D}^{-1}(n),
\end{equation}
we can rewrite Eq. \ref{eq1_7} as:
\begin{displaymath}
 \mathbf{A}(0)=\underbrace{\mathbb{D}^{-1}(0)\mathbb{T}(1)\mathbb{T}(2)\dots \mathbb{T}(N-1)\mathbb{T}(N)\mathbb{D}(s)}_{=\mathbb{T}}\mathbf{A}(S) .
\end{displaymath}
The $4\times 4$ matrix equation (\ref{eq1_8}) yields unique solutions $\mathbf{A}(0)$ and $\mathbf{A}(S)$, provided that four boundary conditions are fixed (see Eq. \ref{eq:A_boundary} below).
In order to obtain the transmission coefficient for the intensity, the Poynting vector for the substrate has to be calculated.
For an isotropic substrate, the time-averaged Poynting vector $\langle\mathbf{S}\rangle$ can be written as:
\begin{equation}\label{eq1_9}
\langle\mathbf{S}\rangle=\frac{1}{2}|\mathbf{p}\times \mathbf{q}|=\frac{|A|^{2}}{\omega\mu_0}|\mathbf{p}\times(\mathbf{k}\times \mathbf{p})|.
\end{equation}
The transmittance in layer $n$ is defined as:
\begin{equation}\label{eq1_10}
\mathcal{T}(n)=\frac{\langle|\mathbf{S}(n)|\rangle}{\langle|\mathbf{S}_0|\rangle}=c \, \mu_0 \, |A(n)|^2 \, |\mathbf{p}(n)\times \mathbf{q}(n)| ,
\end{equation}
where $\mathbf{S}_0$ is the Poynting vector in vacuum.
Since transmittance is typically measured in the substrate layer $S$, we write this coefficient as:
\begin{align}
\mathcal{T}_{p}=c\mu_0 |A_3(S)|^2 |\mathbf{p}_3(S)\times \mathbf{q}_3(S)| \label{eq1_11}, \\ \mathcal{T}_{s}=c\mu_0 |A_1(S)|^2 |\mathbf{p}_1(S)\times \mathbf{q}_1(S)| \label{eq1_112},
\end{align}
where $p$ ($s$) is the parallel (perpendicular) component.
The \textit{total} transmission coefficient is the sum of $p$ and $s$ components: $\mathcal{T}_{tot}=\mathcal{T}_p+\mathcal{T}_s$.
The absorbance $\mathcal{A}$ is directly related to the transmittance through Beer's law:
\begin{equation}\label{eq1_12}
\mathcal{A}=-\ln(\mathcal{T}).
\end{equation}
This relation holds for the $p$ and $s$ components, as well as for the \textit{total} coefficient.
Finally, parallel and perpendicular components of the reflectance $\mathcal{R}_{s,p}$ are expressed, respectively, as:
\begin{equation}\label{eq1_13}
\mathcal{R}_p=|A_4(0)|^{2},
\end{equation}
and
\begin{equation}\label{eq1_14}
\mathcal{R}_s=|A_2(0)|^{2}.
\end{equation}
The \textit{total} reflection coefficient is $\mathcal{R}_{tot}=\mathcal{R}_p+\mathcal{R}_s$. 

\section{Numerical Procedure}\label{sec:implement}
\begin{figure}
\includegraphics[width=.45\textwidth]{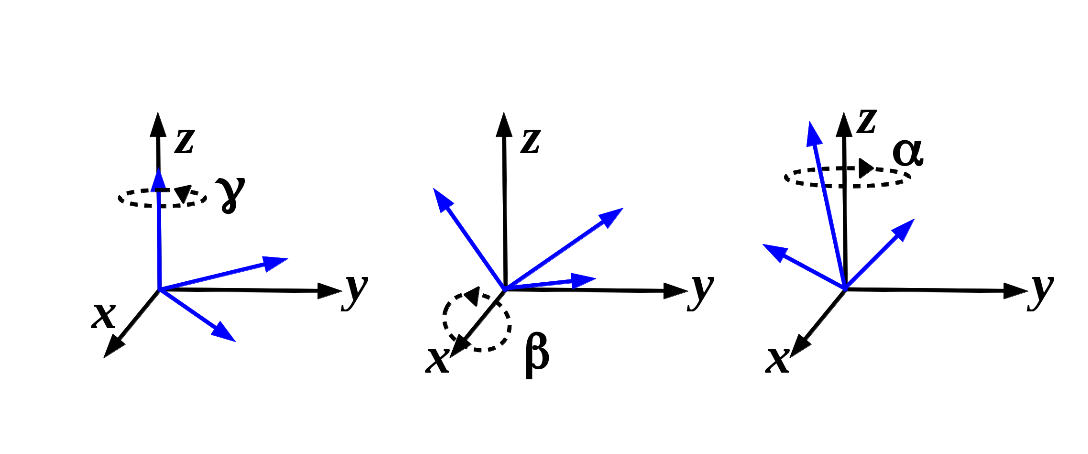}%
\caption{Euler rotations $Z_1X_2Z_3$, as implemented in \texttt{LayerOptics}, for the angles $\alpha$, $\beta$ and $\gamma$ around the Cartesian axes $z$, $x$, and $z$, respectively. Rotations are shown in the order they are applied: first $\gamma$, then $\beta$ and finally $\alpha$. The rotated (fixed) frame is indicated in blue (black). }
\label{fig2:euler}
\end{figure}
%
\subsection{Setup}
In this section, we describe the numerical procedure to compute optical coefficients in layered materials with \texttt{LayerOptics}.
For this purpose, a number of preliminary steps needs to be performed. 
This is reflected in the structure of \texttt{LayerOptics}: Before executing the script, a setup tool has to be run (see Appendix B).
In the beginning, we have to define the notation for the four-component vectors $\mathbf{A}$, with respect to the directions of light propagation in the birifringent medium.
According to the scheme presented in Fig. \ref{fig1:setup}, with the interface between the semi-infinite vacuum layer and the top dielectric layer at $z$=0, we consider \textit{downwards} motion along the ($-z$)-direction, and \textit{upwards} motion in ($+z$)-direction.
Moreover, we define the \textit{parallel} polarization of the electric field in the $zy$-plane, and the \textit{perpendicular} one in the $xz$-plane.
In this way, we can index the four components as follows:
\begin{equation}
\mathbf{A}= \left \{\begin{array}{lll}
 								A_1 & \leftarrow & \text{downwards perpendicular component} \\
 								A_2 & \leftarrow & \text{upwards perpendicular component} \\
 								A_3 & \leftarrow & \text{downwards parallel component} \\
 								A_4 & \leftarrow & \text{upwards perpendicular component}.
 \end{array}\right.
\label{eq:A_boundary}
\end{equation}

Full dielectric tensors, including off-diagonal components, represent the main input of \texttt{LayerOptics}. 
In standard \textit{ab initio} codes dielectric tensors are expressed with respect to Cartesian axes, which in case of non orthogonal unit cells may not coincide with the lattice vectors.
When dealing with anisotropic materials, it is crucial to express the dielectric tensor in terms of a coordinate system that reflects the experimental setup.
For this purpose, we introduce the rotation matrix $\mathbb{R}=Z_1X_2Z_3$, defined by the Euler angles $\alpha$, $\beta$ and $\gamma$ (see Fig. \ref{fig2:euler}).
With these three rotations any possible orientation of the sample with respect to the reference coordinate system can be represented.
Euler rotations are performed in the following order (see Fig. \ref{fig2:euler}): First a rotation $\gamma$ around the $z$-axis is considered, next a rotation $\beta$ is performed around $x$, and finally rotation $\alpha$ with respect to the $z$-axis. 
Using the following standard notation:
\begin{align*}
s_{1}= \sin \alpha, \;\;\;\; c_{1}= \cos \alpha \\
s_{2}= \sin \beta, \;\;\;\; c_{2}= \cos \beta \\
s_{3}= \sin \gamma, \;\;\;\; c_{3}= \cos \gamma
\end{align*}
the rotation matrix is expressed as:
\begin{equation*}\label{eq2_3}
\mathbb{R}=Z_1X_2Z_3=\left(\begin{array}{ccc} c_{1}c_{3}-c_{2}s_{1}s_{3} & -c_{1}s_{3}-c_{2}c_{3}s_{1} & s_{1}s_{2} \\
								   c_{3}s_{1}+c_{1}c_{2}s_{3} & c_{1}c_{2}c_{3}- s_{1}s_{3} & -c_{1}s_{2} \\
								   s_{2}s_{3} & c_{3}s_{2} & c_{2}

\end{array}\right),
\end{equation*}
and the resulting \textit{transformed} dielecric tensor $\epsilon'$ is:
\begin{equation}\label{eq2_4}
\epsilon'=\mathbb{R}\epsilon \mathbb{R}^{-1} .
\end{equation}

Finally, a number of parameters related to the experimental setup has to be chosen (see Fig. \ref{fig1:setup}).
$\Theta$ is the angle between the incident beam, assumed in the $zy$-plane, and the plane of incidence of the sample ($xy$). 
$\delta$ determines the angle of polarization of the incoming light, with amplitude normed to one: $\delta=0$ corresponds to light fully polarized in the parallel direction, while $\delta=\dfrac{\pi}{2}$ indicates incoming light with perpendicular polarization.
In addition to the (optionally rotated) full dielectric tensor $\epsilon(n)$, the thickness $t$ of each layer has to be provided in input. 

\begin{figure}[ht]
\centering
\includegraphics[width=.45\textwidth]{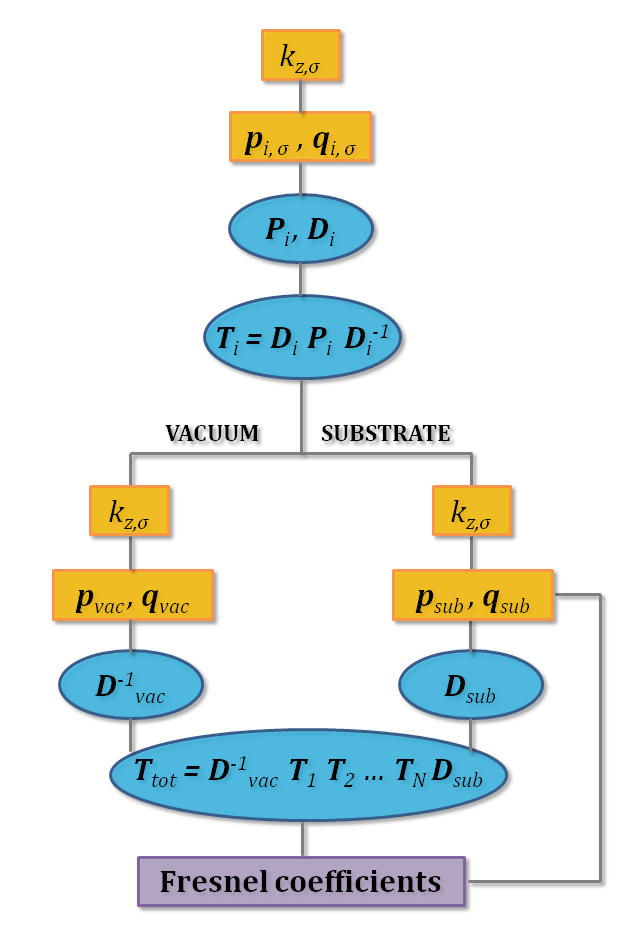}%
\caption{Flowchart of \texttt{LayerOptics}. Rectangles indicate solution of the Maxwell's equations in reciprocal space and elliptic shapes denote matrix construction.}
\label{fig:flowchart}
\end{figure}

\subsection{Calculation of Fresnel coefficients}

Based on the equations presented in Section \ref{sec:theory}, optical coefficients are computed by \texttt{LayerOptics}, following the steps indicated in the flowchart of Fig. \ref{fig:flowchart}.
First, for each layer $i$, the roots of the characteristic polynomial of Eq. \ref{eq1_1} are calculated, yielding the wave-vector components $k_{z,\sigma}$ ($\sigma = 1, ..., 4$). 
From them, the corresponding electric and magnetic polarization vectors, $\mathbf{p}_{\sigma}$ and $\mathbf{q}_{\sigma}$, respectively, are obtained. 
These are the main ingredients to determine the transfer matrices $\mathbb{T}_i$ (Eq.~\ref{eq1_8}), after $\mathbb{D}_i$ and $\mathbb{P}_i$ are calculated from Eqs. \ref{eq1_5} and \ref{eq1_6}.
Special treatment is required for vacuum and substrate layers.
The matrices $\mathbb{D}^{-1}(0)$ and $\mathbb{D}(s)$ are needed in order to determine the total transfer matrix $\mathbb{T}_{tot}$ of the entire layered system.
To compute $\mathbb{T}_{tot}$, Eq. \ref{eq1_8} is solved for $A_1(0)$ and $A_3(0)$, representing the amplitude of the incoming light set in input, and by fixing $A_2(s)=A_4(s)=0$, under the physical assumption that no light is emitted from the substrate, at $z=-\infty$.
In the last step, Fresnel coefficients are calculated, according to Eqs. \ref{eq1_11}-\ref{eq1_112} (transmittance), Eq.~\ref{eq1_12} (absorbance) and Eqs.~\ref{eq1_13}-\ref{eq1_14} (reflectance).

\section{Applications}\label{sec:results}
In this section we present a selection of applications of \texttt{LayerOptics}.
We focus on molecular systems, namely oligothiophene crystals and azobenzene SAMs, where anisotropy may induce pronounced effects.
We apply \texttt{LayerOptics} to optical and X-ray absorption spectra, at varying incidence and polarization angle of the incoming light beam, as well as the orientation of the organic thin film with respect to the substrate.
Dielectric tensors are computed from MBPT, through the solution of the BSE, as implemented in the \texttt{exciting} code \cite{exciting}. 
In all the examples presented below, the substrate is modeled with the frequency-independent dielectric function of bulk silicon ($\epsilon_0$=11.8 \cite{edwa-ocho80ao}).
Since the frequency-independent background is subtracted from the spectra shown in the following, the specific choice of the substrate does not play a role here.
\subsection{Reflection coefficient of sextithiophene thin films}
\begin{figure*}
\includegraphics[width=.95\textwidth]{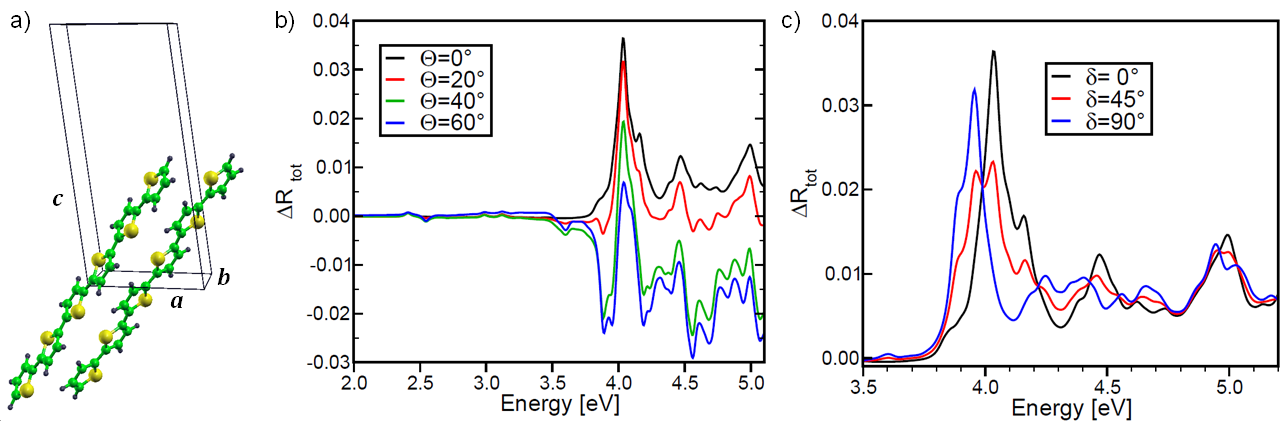}%
\caption{a) Sexithiophene (6T) crystal structure. Deviation of the total reflection coefficient from the background ($\Delta R_{tot}$) at varying angle $\Theta$ of the incoming beam (b), and polarization angle $\delta$ of the incident light (c).}
\label{fig4:polarization}
\end{figure*}
In the first example, we investigate the dependence of the reflection coefficient of a sextithiophene (6T) thin film on the angle and polarization of the incoming light.
In the so-called high temperature phase, 6T is a monoclinic crystal with 2 molecules per unit cell, lattice parameters $a$ = 9.14 \AA{}, $b$ = 5.68 \AA{}, and $c$ = 20.67 \AA{}, and monoclinic angle $\beta$= 97.78$^{\circ}$ between $a$ and $c$ \cite{sieg+95jmr} (see Fig. \ref{fig4:polarization}a).
For this structure, the dielectric tensor has non-zero off-diagonal components $xz$.
We consider a sample of thickness $t$ = 2 nm.

According to the notation in Fig. \ref{fig1:setup}, we analyze 4 configurations by varying the angle of incidence $\Theta$ of the incoming light beam.
In addition to $\Theta=0^{\circ}$, corresponding to normal incidence, we consider $\Theta=20^{\circ}$, $\Theta=40^{\circ}$, and $\Theta=60^{\circ}$.
The deviation of the total reflection coefficient from the frequency-independent background ($\Delta \mathcal{R}_{tot}$) is shown in Fig.~\ref{fig4:polarization}.
From Fig. \ref{fig4:polarization}b, we notice that in the visible region (2 -- 3 eV) $\Delta\mathcal{R}_{tot}$ is very low, almost independently of the value of $\Theta$.
At about 2.5 eV, two bound intramolecular excitons appear, as described in Ref. \cite{pith+15cgd}.
These excitons have weak oscillator strength: hence, the incidence angle of the incoming beam has an almost negligible effect.
On the contrary, at higher energies (3.5 -- 5 eV), more intense excitations characterize the spectrum, and consequently $\Delta \mathcal{R}_{tot}$ undergoes larger variations depending on $\Theta$. 
At normal incidence, $\Delta \mathcal{R}_{tot}$ is always positive, and peaks are observed at 4, 4.5, and 5 eV.
Similar features  are observed for $\Theta=20^{\circ}$, where, however, the shoulders turn into dips, with $\Delta \mathcal{R}_{tot} < 0$.
A different scenario appears for $\Theta=40^{\circ}$ and $\Theta=60^{\circ}$.
In both cases $\Delta \mathcal{R}_{tot}$ is characterized by pronounced dips, and, except for the intense feature at about 4 eV, it is constantly negative in the region 3.5 -- 5 eV.

Next, we consider the dependence of $\Delta \mathcal{R}_{tot}$ on the polarization direction of the incoming light.
For normal incidence, we vary the polarization angle $\delta$ (see Fig. \ref{fig1:setup}) from 0$^{\circ}$ (parallel to the $y$ axis) to 90$^{\circ}$ (parallel to the $x$ axis).
The corresponding plot for $\Delta \mathcal{R}_{tot}$ is shown in Fig. \ref{fig4:polarization}c.
As mentioned above, for $\Theta=0^{\circ}$ $ \Delta\mathcal{R}_{tot}$ is always positive.
For parallel light polarization ($\delta$ = 0$^{\circ}$) we observe again an intense peak at 4 eV, with a shoulder at about 4.3 eV. 
Additional peaks, with lower oscillator strength, appear between 4.5 eV and 5 eV.
In case of perpendicular light polarization ($\delta$ = 90$^{\circ}$), $\Delta \mathcal{R}_{tot}$ exhibits different features.
The most intense peak is now found at $\sim$3.9 eV, with a shoulder at $\sim$3.8 eV.
Also at higher energies, in the region 4.25 -- 4.75 eV, peaks appear for $\delta$ = 90$^{\circ}$ where dips are observed for parallel polarization of the incoming light.
Only the peak at 5 eV is present, regardless the value of $\delta$.
For incoming light with $\delta$=45$^{\circ}$, $\Delta \mathcal{R}_{tot}$ shows a combination of the features observed in the previous cases ($\delta$ = 0$^{\circ}$ and $\delta$ = 90$^{\circ}$).
This is especially evident in the excitation band around 4 eV, where two peaks and two shoulders appear, corresponding to the respective features observed for parallel and perpendicular polarization angles.
The same holds also for the peaks between 4.25 and 4.75 eV.

\subsection{Optical absorption of layered bithiophene thin films}
\begin{figure}
\includegraphics[width=.48\textwidth]{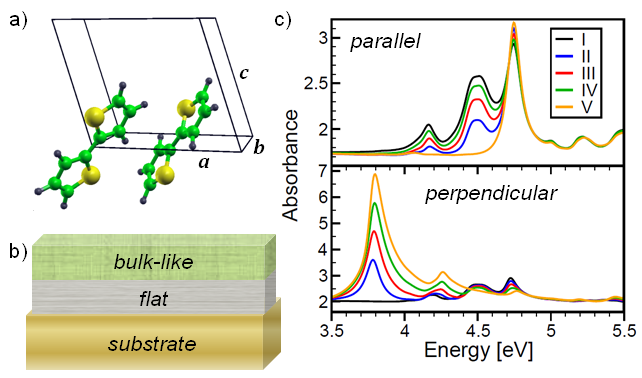}%
\caption{a) Unit cell of a bithiophene (2T) crystal. b) Layered structure of the 2T thin film: in the \textit{bulk-like} layer, the molecules are oriented as in a), with the $ab$ plane of the unit cell at the interface, while in the \textit{flat} configuration the unit cell is rotated such that the long molecular axis is parallel to the substrate surface. c) Parallel and perpendicular components of the absorbance for each setup, ranging from I to V (see Table \ref{table}). The contribution of the frequency-independent substrate is subtracted.}
\label{fig5:2T}
\end{figure}
In the second example we consider a layered thin film of a bithiophene (2T) crystal.
Like 6T, also 2T has a monoclinic unit cell, with lattice parameters $a$ = 7.73 \AA{}, $b$ = 5.73 \AA{}, and $c$ = 8.93 \AA{} and monoclinic angle $\beta$= 106.72$^{\circ}$ between $a$ and $c$ \cite{pell-bris94acc}, hosting 2 molecules (see Fig. \ref{fig5:2T}a).
We consider a system with total thickness $t_{tot}$ = 20 nm, consisting of two layers of 2T.
Each layer is characterized by a different orientation of the molecules with respect to the substrate, as sketched in Fig. \ref{fig5:2T}b.
In the lower layer, at the boundary with the isotropic substrate, the molecules lie \textit{flat}, with their long axis oriented parallel to the surface.
This is typically the situation of organic thin films grown on a metal substrate (see e.g. Ref. \cite{loi+05natm}).
Such configuration is obtained by applying a rotation (Fig. \ref{fig5:2T}a) of -52.4$^{\circ}$ around the $y$ axis of the unit cell.
In the upper layer, the 2T thin film lies in the $ab$ plane of its unit cell: We refer to this as \textit{bulk-like} configuration.
No Euler rotation is applied in this case.

\begin{table}
\centering
 \begin{tabular}{ccc}
\hline
\hline
Setup  & $t_{flat}$ & $t_{bulk-like}$ \\
\hline
I & 20 & 0 \\
II & 15 & 5 \\
III & 10 & 10 \\
IV & 5 & 15 \\
V & 0 & 20 \\
 \hline
 \hline
\end{tabular}
\caption{Different setups of a layered 2T thin film, with total thickness of 20 nm. The thickness $t_{flat}$ ($t_{bulk-like}$), in nm, is referred to the \textit{flat} (\textit{bulk-like}) orientation of the molecules with respect to the substrate (see Fig. \ref{fig5:2T}a-b).}
\label{table}
\end{table}

We consider overall 5 setups, keeping $t_{tot}$ = 20 nm fixed, and varying the thickness of the \textit{flat} and \textit{bulk-like} layers with steps of 5 nm (see Table \ref{table}).
For each setup, we compute parallel ($p$) and perpendicular ($s$) components of the absorbance, shown in Fig. \ref{fig5:2T}c.
Again, we model the substrate using the frequency-independent dielectric constant of bulk silicon, and we subtract the background. 
By comparing the set of spectra for $p$ and $s$ polarization, we immediately notice striking differences.
The parallel component of the absorbance presents a maximum at about 4.75 eV, which appears for each setup with almost the same intensity.
At lower energy, at approximately 4.2 and 4.5 eV, respectively, two additional features appear in the $p$-polarized spectrum of those configurations incorporating a \textit{flat} layer.
When all molecules in the thin film are oriented to lie \textit{flat} onto the substrate (setup I), these peaks disappear and only an extremely weak shoulder is present just above 4 eV.
Above 5 eV, all the spectra coincide.
Conversely, in the $s$-polarized component of the absorbance, the maximum is found at about 3.8 eV, and again it is most intense in the pure \textit{bulk-like} configuration (V).
The strength of this peak decreases at increasing thickness of the \textit{flat} layer, and completely disappears in setup I (\textit{flat} configuration only).
Above 4 eV, weaker features are observed.
The shoulder at about 4.25 eV and the weak peak at $\sim$4.8 eV behave similarly to the most intense peak, i.e., their intensity is maximum (minimum) in setup V (I). 
On the other hand, the peak at 4.5 eV maintains almost the same intensity going from system I to IV, while it is not present in the pure \textit{bulk-like} configuration (setup V).

\subsection{Polarization-resolved X-ray absorption spectra of azobenzene-functionalized self-assembled monolayers}
\begin{figure}
\centering
\includegraphics[width=.45\textwidth]{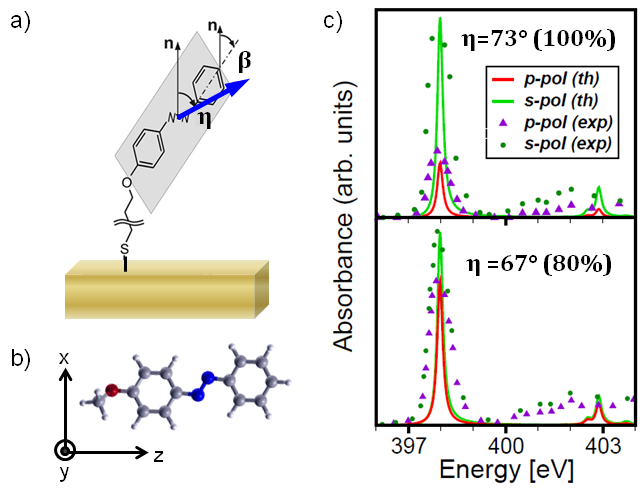}%
\caption{a) Schematic representation of an azobenzene-functionalized SAM of alkanethiols on a substrate. The mean tilt angle $\eta$ is defined between the normal (blue arrow) of the phenyl rings plane (shaded rectangle) and the normal with respect to the substrate ($\mathbf{n}$ -- black arrow). $\beta$ is the angle between $\mathbf{n}$ and the long molecular axis. b) Azobenzene molecule considered in the calculation of dielectric tensors: an \ce{O-CH3} end group terminates the molecules to simulate the chemical environment of the covalent bond to the SAM. c) Parallel ($p$) and perpendicular ($s$) components of the absorbance at varying $\alpha$, corresponding to different concentration of azobenzene molecule in the SAM. Experimental data are taken from Ref. \cite{mold+15lang}.}
\label{fig6:azobenzene}
\end{figure}
As a final example, we present the application of \texttt{LayerOptics} to X-ray absorption spectra (XAS) from the nitrogen (N) K-edge of azobenzene-functionalized SAMs.
In a recent experiment \cite{mold+15lang}, it was shown that the orientation of the azobenzene molecules with respect to the substrate depends on their concentration in the SAM. 
Such behavior can be observed from polarization-resolved XAS, where the intensity of the peaks, and in particular of the lowest-energy resonance, varies significantly between parallel and perpendicular components for different azobenzene concentrations.
This resonance corresponds to a transition from N 1$s$ to the LUMO, which has $\pi^*$ character and transition dipole moment along the long molecular axis \cite{mold+15lang,cocc-drax15condmat}.
For this reason, its strength is expected to be affected by the orientation of the molecule with respect to the substrate.
This can be determined by estimating the mean tilt angle $\eta$ (see Fig. \ref{fig6:azobenzene}a), which represents the angle between the normal of the phenyl rings plane (thick blue arrow) and the normal to the substrate ($\mathbf{n}$, black arrow).

Our starting point is the dielectric tensor of the azobenzene molecule, including excitations from N 1$s$ core levels to the conduction states.
The molecule has been accommodated in an orthorhombic supercell with lattice parameters oriented according to the coordinate axes in Fig. \ref{fig6:azobenzene}b (for additional details, see Ref. \cite{cocc-drax15condmat}).
In these calculations, we have neglected the presence of the alkyl chains, which connect azobenzene to the gold substrate in the experimental sample (see Fig. \ref{fig6:azobenzene}a), since, for core-level excitations from the N K-edge, they are expected not to play a role.
We represent the SAM as a thin film of thickness 4 nm, and we model the substrate using the frequency-independent dielectric function of bulk silicon \bibnote{Although experimentally a gold substrate is used, we model it with silicon for simplicity. We expect this choice to have no impact on our results.}.
The angle of incidence of incoming light is set to $\Theta$=70$^{\circ}$, as in experiment.
We compute the $p$- and $s$-components of the absorbance at different values of $\eta$, corresponding to azobenzene concentration of 100$\%$ ($\eta$=73$^{\circ}$) and 80$\%$ ($\eta$=67$^{\circ}$) \cite{mold+15lang}, \bibnote{The angles of azobenzene molecules in a SAM with concentration of 80$\%$ are predicted from experiment to range between 54$^{\circ}$ and 64$^{\circ}$. In our simulation, we consider a value of $\eta$=67$^{\circ}$, which is slightly above the experimental prediction. Due to the large uncertainty also in experiment, we consider this choice fully acceptable.}.
Different orientations of the molecules with respect to the substrate are represented by rotations of the dielectric tensor.
The Euler angle $\beta$, between the normal direction with respect to the substrate $\mathbf{n}$ and the long molecular axis (see Fig. \ref{fig6:azobenzene}a), is fixed to 30$^{\circ}$.
Under this condition, $\eta$ is related to the third Euler angle $\gamma$ by $\cos \eta=\frac{\cos \gamma}{2}$.
Making use of this relationship and setting the first Euler angle $\alpha$=0$^{\circ}$, we can effectively simulate different molecular orientations indicated by the angle $\eta$, by rotating the dielectric tensor with respect to $\gamma$: $\eta$=73$^{\circ}$ corresponds to $\gamma$=55$^{\circ}$ and $\eta$=67$^{\circ}$ to $\gamma$=40$^{\circ}$.
The resulting absorption spectra are shown in Fig. \ref{fig6:azobenzene}c. 
Theoretical and experimental data for the absorbance are normalized to the height of the first resonance \bibnote{Experimental data are given by Auger yield in arbitrary units (see Ref. \cite{mold+15lang})}.
The absorbance in the top panel, corresponding to $\eta$=73$^{\circ}$, i.e., full azobenzene coverage in the SAM, shows a rather strong polarization-dependence of the intensity of the first resonance, which is at least twice more intense in the $s$-component, compared to the parallel one.
By decreasing $\eta$ to 67$^{\circ}$, as predicted in the SAM with 80$\%$ azobenzene concentration, the relative spectral weight of the $p$-component increases for the first peak.
Although the peak in the $s$-component remains stronger, its relative weight with respect to the parallel component is significantly decreased.
As shown in Fig. \ref{fig6:azobenzene}c, the relative height of the peaks is in good agreement with the experimental data in both cases. 
Also at higher energy, the oscillator strength of the peaks at about 403 eV is matched well by our calculations \bibnote{The slight blue-shift of our results compared to the experimental spectrum are ascribed to the use of the dielectric tensor the the isolated molecule, while the data from Ref. \cite{mold+15lang} are taken for the SAM. For an exhaustive analysis and comparison of the XAS of azobenzene molecule and SAM see Ref. \cite{cocc-drax15condmat}.}.

Finally, with \texttt{LayerOptics}, we are also able to estimate the so-called \textit{magic angle}, corresponding to the polarization angle for which the absorbance is independent of the polarization channel $s$ or $p$.
Experimentally, it is considered at 54.7$^{\circ}$ \cite{mold+15lang}.
We can determine this angle by tuning the polarization angle such that the parallel component of the absorbance ($\mathcal{A}_{p}$) is equivalent to the perpendicular one ($\mathcal{A}_{s}$).
From our results, this occurs at 59.6$^{\circ}$.
Considering that experimentally these angles are given with an error bar of $\pm$5$^{\circ}$ \cite{mold+15lang}, we regard this estimate as satisfactory.

\section{Summary and Conclusions}
We have presented \texttt{LayerOptics}, an implementation of the 4$\times$4 matrix formalism of Maxwell's equations to compute Fresnel coefficients in anisotropic layered thin films.
We are able to match the laboratory setup, by taking into account the direction and the polarization of the incoming light, as well as the orientation of each layer with respect to the substrate.
The capabilities of \texttt{LayerOptics} have been demonstrated by two different scenarios, such as optical properties of organic thin films on a substrate, and polarization-dependent X-ray absorption spectra of azobenzene-functionalized SAMs.
With prototypical examples, we have shown the importance of off-diagonal components of dielectric tensors in anisotropic materials, in order to quantitatively compare the calculated spectra with experiments.
Incorporating information about orientation and polarization of the light allows for an even more quantitative comparison.
Our results confirm the potential of this approach to gain insight into the microscopic mechanisms ruling light absorption in anisotropic thin films.  
\texttt{LayerOptics} is applied to post-process full dielectric tensors computed with the \texttt{exciting} code.
Due to its simple and versatile structure, this tool can be easily interfaced to any other \textit{ab initio} package.

\section*{Acknowledgement}
This work was funded by the German Research Foundation (DFG), through the Collaborative Research Centers SFB 658 and 951. C.C.~acknowledges support from the \textit{Berliner ChancengleichheitsProgramm} (BCP).
We are grateful to Juan Pablo Echeverry Enciso for providing the dielectric tensors of the 2T crystal.

\section*{Appendix A: Solutions of Maxwell's equations for strongly anisotropic materials}
Among the four solutions $k_{z,\sigma}$ of the Maxwell's wave equation (Eq. \ref{eq1_1}), it is commonly assumed that two have positive real part ($\Re(k_{z,\sigma})>0$), while for the other two $\Re(k_{z,\sigma})<0$. 
This corresponds to the physical situation of electromagnetic waves moving through the medium up- and down-wards, with respect to the incidence direction of the light beam \cite{Yeh}.
We find, however, that this assumption is not generally fulfilled when the off-diagonal elements of the dielectric tensors $\epsilon$ are of the same order of magnitude of the diagonal ones.
This can happen in strongly anisotropic materials.
In this case, $\Re(k_{z,\sigma})$ may have the same sign in all four components of $\epsilon$, giving rise, at the same time, to non-zero transmission and reflection coefficients in the layered system.
This occurrence leads to an arbitrariness in the notion of \textit{parallel} and \textit{perpendicular} components with respect to the plane of incidence of the incoming beam, as well as in the notion of \textit{upwards} and \textit{downwards} directions of light propagation with respect to the layer. 
Since \texttt{LayerOptics} assumes the electric amplitudes of the vacuum and the substrate layers to be isotropic, the boundary condition that no light is emitted from $z=-\infty$ is fulfilled.
Hence, the physical meaning of the computed Fresnel coefficients is not affected.


\section*{Appendix B: Input and Output}
\label{appendix:IO}
\begin{figure}
\fbox{\includegraphics[width=.45\textwidth]{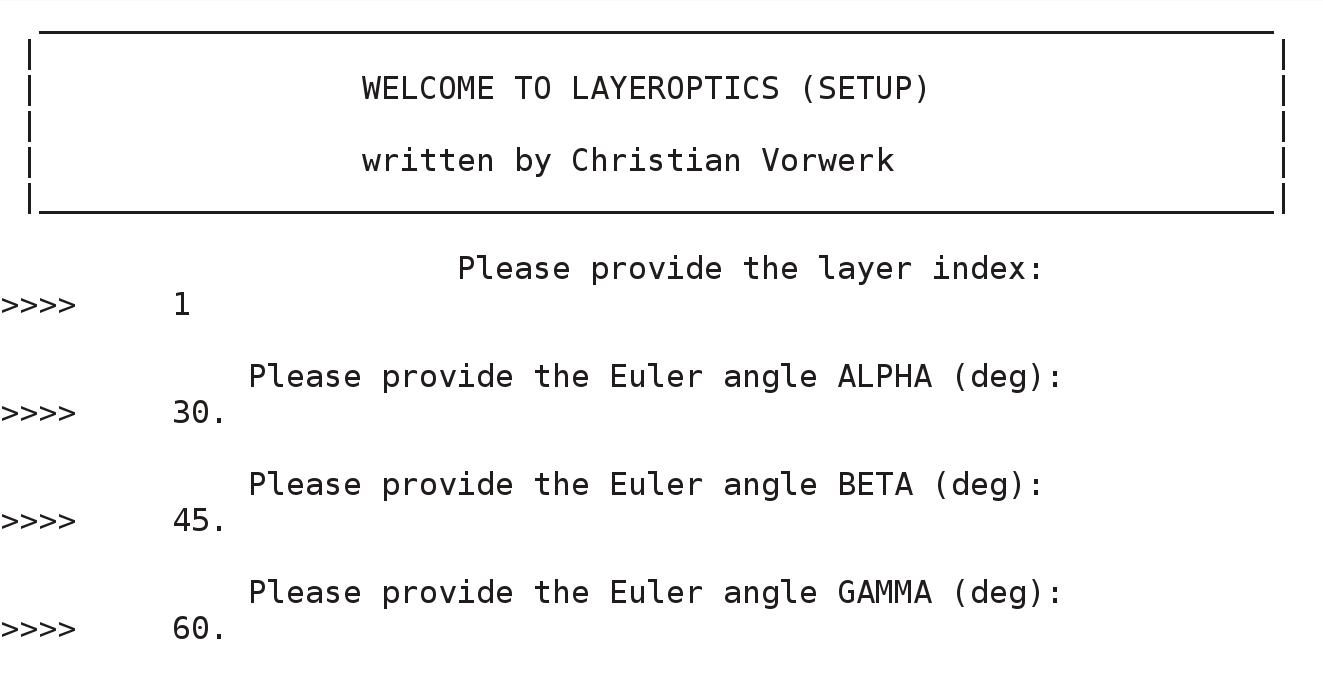}}
\caption{Interactive interface of the script \texttt{LO-setup.py}. In this example, the layer index is 1, and a Euler rotation of the dielectric tensor is set according to the angles $\alpha$=30$^{\circ}$, $\beta$=45$^{\circ}$, and $\gamma$=60$^{\circ}$.}
\label{fig:LOsetup}
\end{figure}

The main input of \texttt{LayerOptics} is given by full dielectric tensors $\epsilon$, including diagonal and off-diagonal components, for each computed energy point.
If any component is missing in the input files, the program automatically sets it to zero for each frequency point.
\texttt{LayerOptics} is a python script, which in the current version works as post-processing tool of the \texttt{exciting} code \cite{exciting}.
Through an interactive interface, the script \texttt{LO-setup.py} allows to label the dielectric tensors for each layer and to transform them using Euler rotations.
The user is asked to define the Euler angles $\alpha$, $\beta$ and $\gamma$, according to the framework depicted in Fig. \ref{fig2:euler}.
In the example shown in Fig. \ref{fig:LOsetup}, the dielectric tensor of layer 1 is rotated according to Euler angles $\alpha$=30$^{\circ}$, $\beta$=45$^{\circ}$, and $\gamma$=60$^{\circ}$.
The transformed tensors are renamed as \texttt{n\_ij.OUT}, where \texttt{n} indicates the layer index, and \texttt{ij} the components of $\epsilon_{ij}$.
\begin{figure}
\fbox{\includegraphics[width=.45\textwidth]{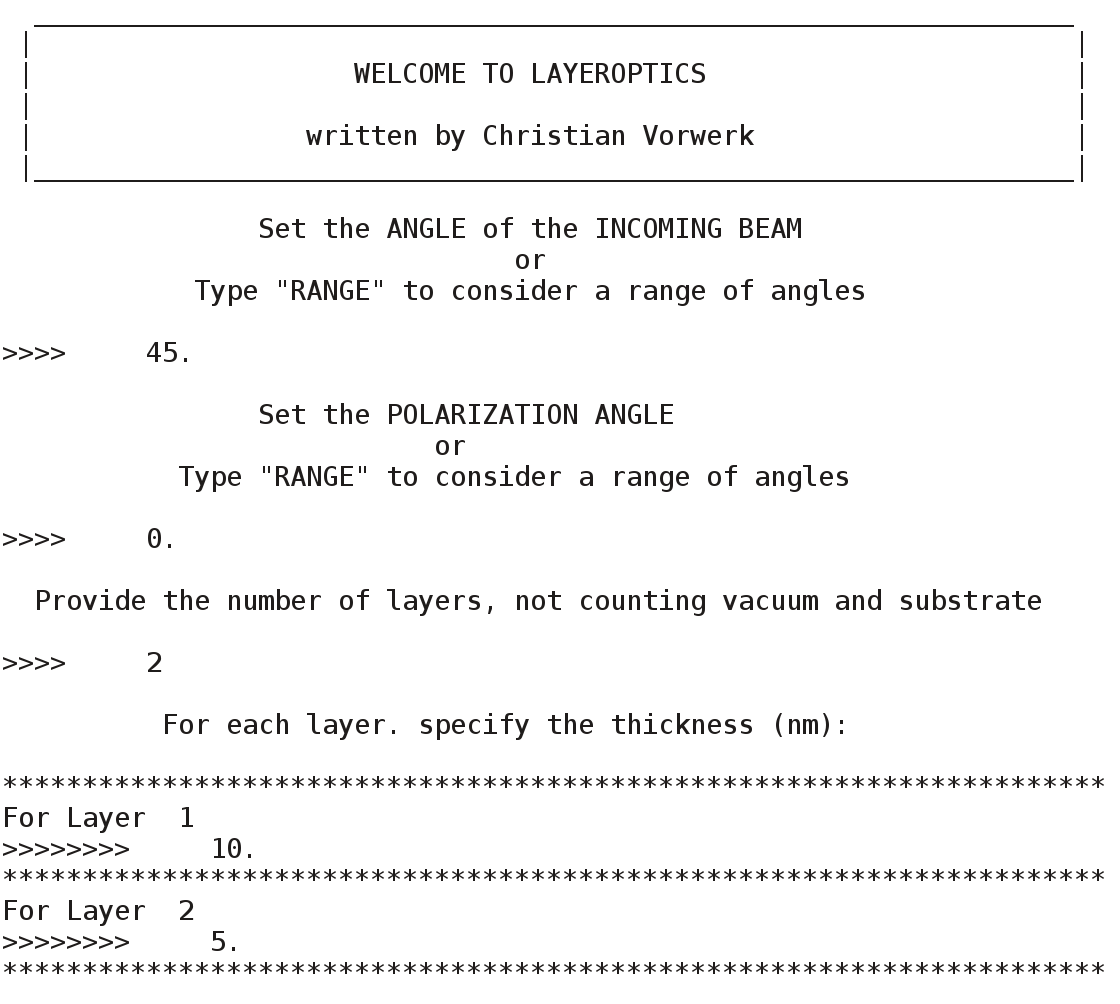}}
\caption{Interactive interface of the script \texttt{LO-execute.py}. In this example, the incidence angle $\Theta$ is set to $45^{\circ}$, and the polarization angle $\delta$ to $0^{\circ}$. The presented system is composed of 2 layers, the first of thickness 10 nm, the second one of thickness 5 nm.}
\label{fig:LOexec}
\end{figure}
These files are used by the script \texttt{LO-execute.py}, which implements the algorithm for calculating Fresnel coefficients (see Fig. \ref{fig:LOexec}).
Also in this case, an interactive interface allows the user to set the input parameters for the incoming light and the number of layers, excluding vacuum and substrate.
The value of the frequency-independent dielectric function for the substrate only determines the background of the optical coefficients, but has no impact on their spectral shape.
The incidence and polarization angles, $\Theta$ and $\delta$, respectively, can be either set by single values or through a range.
In the latter case, the user has to specify the initial and final value of the interval, as well as the number of intermediate steps.
The user is also asked to set the thickness of each layer.
An example for a 2-layer sample is shown in Fig. \ref{fig:LOexec}.
The output consists of three files: \texttt{absorbance.out}, \texttt{reflection.out} and \texttt{transmission.out}.
Each file contains 4 columns: energy (in eV), parallel ($p$) component, perpendicular ($s$) component, and total value of the corresponding Fresnel coefficient.
For calculations performed over a range of $\Theta$ or $\delta$ angles, for each point of the interval a separate file is produced, labeled by a number, starting with zero (e.g. \texttt{absorbance0.out}).


\end{document}